\documentclass[12pt]{article}

\usepackage{amsmath,amsfonts,amssymb}

\usepackage{subfigure}

\usepackage{epsfig,graphicx,bm}

\usepackage{epstopdf}

\usepackage{booktabs}

\usepackage{lipsum}

\begin{document}
	
	\begin{titlepage}
		
		\begin{center}
			
			\vskip 0.4 cm

				{\Large \bf  BBGKY Hierarrchy for $N$ D0-Branes}

			\vskip 1cm
			
			\vspace{1em}J. Kluso\v{n},		
			\footnote{Email addresses:	 klu@physics.muni.cz   (J.Kluso\v{n}), }\\
			
			\vspace{1em}
\textit{Department of Theoretical Physics and
				Astrophysics, Faculty of Science,\\
				Masaryk University, Kotl\'a\v{r}sk\'a 2, 611 37, Brno, Czech
				Republic}
			
			\vskip 0.8cm
			
		\end{center}
		
		\begin{abstract}
We study statistical description of   $N$ D0-branes system that is defined by matrix mechanics.
 We determine  BBGKY hierarchy for collection of distribution 
functions $\rho_n$ that gives exact statistical description of this system.

		\end{abstract}
	\end{titlepage}
	
	\bigskip
	
	\newpage

\def\bn{\mathbf{n}}
\newcommand{\bC}{\mathbf{C}}
\newcommand{\bD}{\mathbf{D}}
\def\hf{\hat{f}}
\def\tK{\tilde{K}}
\def\mC{\mathcal{C}}
\def\bJ{\mathbf{J}}
\def\bL{\mathbf{L}}
\def\bk{\mathbf{k}}
\def\tr{\mathrm{tr}\, }
\def\tmH{\tilde{\mH}}
\def\tY{\mathcal{Y}}
\def\nn{\nonumber \\}
\def\bI{\mathbf{I}}
\def\tmV{\tilde{\mV}}
\def\e{\mathrm{e}}
\def\bP{\mathbf{P}}
\def\bE{\mathbf{E}}
\def\bX{\mathbf{X}}
\def\bZ{\mathbf{Z}}
\def\bY{\mathbf{Y}}
\def\bR{\bar{R}}
\def\hN{\hat{N}}
\def\bphi{\mathbf{\Phi}}
\def\hK{\hat{K}}
\def\hnabla{\hat{\nabla}}
\def\hc{\hat{c}}
\def\mH{\mathcal{H}}
\def \Gi{\left(G^{-1}\right)}
\def\hZ{\hat{Z}}
\def\bz{\mathbf{z}}
\def\bK{\mathbf{K}}
\def\iD{\left(D^{-1}\right)}
\def\tmJ{\tilde{\mathcal{J}}}
\def\tr{\mathrm{Tr}}
\def\mJ{\mathcal{J}}
\def\partt{\partial_t}
\def\parts{\partial_\sigma}
\def\bG{\mathbf{G}}
\def\str{\mathrm{Str}}
\def\Pf{\mathrm{Pf}}
\def\bM{\mathbf{M}}
\def\tA{\tilde{A}}
\newcommand{\mW}{\mathcal{W}}
\def\bx{\mathbf{x}}
\def\by{\mathbf{y}}
\def \mD{\mathcal{D}}
\newcommand{\tZ}{\tilde{Z}}
\newcommand{\tW}{\tilde{W}}
\newcommand{\tmD}{\tilde{\mathcal{D}}}
\newcommand{\tN}{\tilde{N}}
\newcommand{\hC}{\hat{C}}
\newcommand{\hg}{g}
\newcommand{\hX}{\hat{X}}
\newcommand{\bQ}{\mathbf{Q}}
\newcommand{\hd}{\hat{d}}
\newcommand{\tX}{\tilde{X}}
\newcommand{\calg}{\mathcal{G}}
\newcommand{\calgi}{\left(\calg^{-1}\right)}
\newcommand{\hsigma}{\hat{\sigma}}
\newcommand{\hx}{\hat{x}}
\newcommand{\tchi}{\tilde{\chi}}
\newcommand{\mA}{\mathcal{A}}
\newcommand{\ha}{\hat{a}}
\newcommand{\tB}{\tilde{B}}
\newcommand{\hrho}{\hat{\rho}}
\newcommand{\hh}{\hat{h}}
\newcommand{\homega}{\hat{\omega}}
\newcommand{\mK}{\mathcal{K}}
\newcommand{\hmK}{\hat{\mK}}
\newcommand{\hA}{\hat{A}}
\newcommand{\mF}{\mathcal{F}}
\newcommand{\hmF}{\hat{\mF}}
\newcommand{\tk}{\tilde{k}}
\newcommand{\hQ}{\hat{Q}}
\newcommand{\mU}{\mathcal{U}}
\newcommand{\hPhi}{\hat{\Phi}}
\newcommand{\hPi}{\hat{\Pi}}
\newcommand{\hD}{\hat{D}}
\newcommand{\hb}{\hat{b}}
\def\I{\mathbf{I}}
\def\tW{\tilde{W}}
\newcommand{\tD}{\tilde{D}}
\newcommand{\mG}{\mathcal{G}}
\def\IT{\I_{\Phi,\Phi',T}}
\def \cit{\IT^{\dag}}
\newcommand{\hk}{\hat{k}}
\def \cdt{\overline{\tilde{D}T}}
\def \dt{\tilde{D}T}
\def\bra #1{\left<#1\right|}
\def\ket #1{\left|#1\right>}
\def\mV{\mathcal{V}}
\def\Xn #1{X^{(#1)}}
\newcommand{\Xni}[2] {X^{(#1)#2}}
\newcommand{\bAn}[1] {\mathbf{A}^{(#1)}}
\def \bAi{\left(\mathbf{A}^{-1}\right)}
\newcommand{\bAni}[1]
{\left(\mathbf{A}_{(#1)}^{-1}\right)}
\def \bA{\mathbf{A}}
\newcommand{\bT}{\mathbf{T}}
\def\bmR{\bar{\mR}}
\newcommand{\mL}{\mathcal{L}}
\newcommand{\mbQ}{\mathbf{Q}}
\def\mat{\tilde{\mathbf{a}}}
\def\mtF{\tilde{\mathcal{F}}}
\def \tZ{\tilde{Z}}
\def\mtC{\tilde{C}}
\def \tY{\tilde{Y}}
\def\pb #1{\left\{#1\right\}}
\newcommand{\E}[3]{E_{(#1)#2}^{ \quad #3}}
\newcommand{\p}[1]{p_{(#1)}}
\newcommand{\hEn}[3]{\hat{E}_{(#1)#2}^{ \quad #3}}
\def\mbPhi{\mathbf{\Phi}}
\def\tg{\tilde{g}}
\newcommand{\phys}{\mathrm{phys}}

\section{Introduction and Summary}
Kinetic theory is one of the cornerstones of theoretical physics with broad areas of applications. It  allows derivation 
of hydrodynamics equations from the first principles based on the Boltzman's idea of microscopic structure
of fluids
\footnote{For extensive review, see \cite{kintheory}.}. Further, it is useful for description of some phenomena in plasma physics
and fusion \cite{Chen}, high-energy particle physics experiments
like  heavy ion collisions \cite{Busza:2018rrf}  and cosmology \cite{Galaxy}.
Kinetic theory  is consistently described with the help of so-called  BBGKY (Bogoliubov-Born-Green-Kirkwood-Yvon) hierarchy 
\cite{Yvon,Bogoliubov,Kirkwood,Born}
that provides exact description of $N$-particle system with the help of distribution functions.
Even if the BBGKY hierarchy is theoretically complete it is very difficult to solve it without
approximations due to the fact that for $N$ that in typical situations is large, the number of equations
is huge. Then various approximations have to be imposed when we want to find kinetic equation
which is an equation for one particle distribution function \cite{kintheory}. Despite of this fact 
BBGKY hierarchy is very powerful method in temporary theoretical physics as well as can be seen 
in recent works \cite{Grozdanov:2024fxr,Bertini:2022zyq,Kaniadakis:2003gbo}.

In this paper we would like to find BBGKY hierarchy for  description  
of  very interesting non-relativistic system which is low energy Lagrangian for $N$ D0-branes in string theory 
\cite{Polchinski:1995mt,Dai:1989ua}. It is well known that 
the low-energy Lagrangian for a system of many type IIA D0-branes is the matrix
quantum mechanics Lagrangian arising from the dimensional reduction to $0 + 1$ dimensions
of the $10$ dimensional super Yang-Mills Lagrangian, for review see 
for example  \cite{Sen:1998kr,Taylor:1997dy,Taylor:1999qk}. An importance of this action is that it is the key point in the formulation of Matrix theory \cite{Banks:1996vh} (based on the earlier work 
 \cite{deWit:1988wri}) which is strictly speaking defined for infinite number of D0-branes even if there is a version of Matrix theory that is correct for finite $N$ as well \cite{Seiberg:1997ad,Sen:1997we}. Generally this system possesses
 $N^2\times 9$ coordinates $\Phi^I_{ab},I=1,2,\dots,9$ and conjugate momenta $\Pi_{I,ab}$ so that the phase space is
 $2\times N^2\times 9$ dimensional which implies that it is very huge in case of large $N$. Then it is natural to perform step from the classical Hamiltonian description when the dynamics of the system is represented by the point in the phase space to the continuous description with the help of distribution function 
 $\rho_N(\Phi,\Pi)$ that describes probability that the system is in the point of space $\Phi_{ab},
\Pi_{ab}$. Now since configuration of $N$ D0-branes is well defined Hamiltonian system Liouville theorem for $N^2$ distribution function is valid. However our goal is to find equation of motion for reduced distribution function $\rho_n$ that describes probability, that phase space variables of submatrix of dimension $n\times n$ are at the specific subspace of phase space. Such a reduced distribution function is defined when ignore information about remaining degrees of freedom. In other words, since $\rho_N$ is probability density we define $\rho_n$ by appropriate integration over complement of the subspace of the phase space whose precise definition will be given bellow. In the same way we split the total Hamiltonian for $N$-D0-branes into the part that describes solely first $n-$D0-branes and the part that describes remaining degrees of freedom. With such a form of Hamiltonian we can now finally proceed to the analysis of  time evolution of reduced distribution function following \cite{kintheory}. If we demand that distribution function vanishes
for asymptotic values of coordinates and momenta we obtain that many terms in the equations of motion for restricted distribution function vanish as a consequence of  integration by parts. Then in the end we find equation of motion for $\rho_n$ 
that on the left side has the form of standard $n-$D0-brane Liouville operator while the right side has still rather complicated form. Finally imposing natural condition that $\rho_N$ is symmetric in its arguments which physically mean that all matrix degrees of freedom are equivalent we find after an appropriate manipulation that the right side can be expressed as
integro-differential operator acting on $\rho_{n+1}$ distribution function for $(n+1)\times (n+1)$ matrix degrees of freedom. In other words we derived famous BBGKY hierarchy for matrix quantum mechanics which is the main result of this paper. 

This paper should be extended in many directions. The most important one would be to find reasonable arguments how to cut BBGKY hierarchy to get kinetic equation for D0-brane distribution function. This would be extremely interesting since we could then study D0-brane hydrodynamics and also  black hole that could be described with specific matrix model \cite{Maldacena:2023acv}. It would be also very interesting to extend this analysis to supersymmetric matrix mechanics, for recent review, see \cite{Lin:2025iir,Fliss:2025omb}. We hope to return to these problems in future. 

This paper is organized as follows. In the next section (\ref{second}) we introduce Lagrangian for $N$ D0-branes and 
determine its canonical form. Then in section (\ref{second}) 
we introduce distribution function on phase space, define reduced distribution function and develop BBGKY hierarchy. 

 \section{Review of Lagrangian and Hamiltonian Formulations of D0-Branes}
 \label{second}
 We start with the Lagrangian for $N$ D0-branes that at the leading order corresponds to 
 $U(N)$ Super-Yang-Mills mechanical system that has the form
\begin{equation}\label{LD0}
S=\int dt L \ , \quad 
L=\frac{1}{2g_sl_s}\tr\left[\dot{\Phi}^I
\dot{\Phi}_I+\frac{1}{2}[\Phi^I,\Phi^J]
[\Phi_I,\Phi_J]+\mathrm{fermions}\right] \ , 
\end{equation}
where $\Phi^I_{AB} \ , \Phi_I\equiv \delta_{IJ}\Phi^J$ are $N\times N$ Hermitean matrices where $I,J=1,2,\dots,9$ and where $A,B,\dots=1,\dots,N$. Further, $g_s$ is string coupling constant and $l_s$ is the string length. Note that the transverse space is nine-dimensional Euclidean space and repeated indices mean summation over them. Finally this Lagrangian contains terms with fermions that makes the Lagrangian $N=16$ Super-Yang-Mills mechanics. In more details, the Lagrangian (\ref{LD0}) can be defined 
by dimensional reduction of $10$ dimensional $\mathcal{N}=1$ super-Yang-Mills theory 
with gauge group $U(N)$ to the $0+1$ dimensions.
 In what follows we will consider simpler model where we do not study contribution of fermions and we also  ignore gauge field. Finally we tread all matrix elements $\Phi^I_{AB}$ as independent, in other words, we do not presume that these matrices are Hermitean. 

\subsection{Hamiltonian Formalism}
In this section we find Hamiltonian from the Lagrangian (\ref{LD0}). 
In the first step we introduce conjugate momenta to the matrix elements $\Phi^I_{AB}$. From (\ref{LD0}) we obtain
\begin{eqnarray}\label{Piij}
&&	\Pi_{I,AB}=\frac{\delta L}{\delta \dot{\Phi}^I_{AB}}=
	\frac{1}{g_s l_s}\delta_{IJ}(\dot{\Phi}^J)_{BA}
\end{eqnarray}
and let us define Hamiltonian in the standard way
\begin{eqnarray}
&&H_N=\Pi_{I,AB}\dot{\Phi}^I_{AB}-L= \frac{1}{2g_sl_s}\dot{\Phi}^I_{AB}
\delta_{IJ}\dot{\Phi}^J_{BA}-\frac{1}{4g_sl_s}\tr[\Phi^I,\Phi^J]
\delta_{IK}\delta_{JL}[\Phi^K,\Phi^L] \ ,
\nonumber \\
\end{eqnarray}
where we again stress that $\Pi_{I,AB}\dot{\Phi}^I_{AB}$ mean summation over $A,B$ indices. Note that in this paper we will not make 
 distinction between upper and lower indices.
We further  introduce
canonical Poisson brackets
\begin{equation}
	\pb{\Phi^I_{AB},
		\Pi_{J,CD}}=\delta^I_J\delta_{AC}\delta_{BD} \ . 
\end{equation}
Finally we return to the Hamiltonian and express it in the form of canonical variables.
Using (\ref{Piij}) we obtain
\begin{eqnarray}
&&\Pi_{I,AB}\delta^{IJ}\Pi_{J,BA}
=\frac{1}{g_s^2l_s^2}\tr \dot{\Phi}^I\delta_{IJ}\dot{\Phi}^J \nonumber \\
\end{eqnarray}
and  we find that the  Hamiltonian for $N$ D0-branes is equal to
\begin{eqnarray}\label{HN}
H_N=\frac{g_sl_s}{2}\tr \Pi_I\delta^{IJ}\Pi_J-\frac{1}{4g_sl_s}\tr [\Phi^I,\Phi^J]\delta_{IK}\delta_{JL}[\Phi^K,\Phi^L] \ . 
\end{eqnarray}
In the next section we develop statistical description of the system of $N$ D0-branes.

\section{Distribution Function  for $N$ D0-Branes}\label{third}
In this section we propose statistical  description of the system of $N$  D0-branes whose canonical Hamiltonian is given in (\ref{HN}). Recall that $\Phi^I$ is real matrix so that we have  $2\times 9\times N^2$ canonical variables  $\Phi^I_{AB}$ with conjugate momenta $
\Pi_{I,AB}$. Let us introduce $18\times N^2$ dimensional phase space which is spanned by these variables. Note that each microstate of  system is represented by a point in the phase space and that corresponds to 
exact configurations of all phase-space variables. In case when the number of degrees of freedom is huge then this picture, even if it is exact, is not very useful and it is more natural to introduce description of this system  in terms of statistical physics. In more details, let us introduce $\mathcal{N}$ identical copies of given system that differ by initial conditions. Collection of these hypothetical systems is called as \emph{Ensemble}. Now let us presume that members of this system continuously fill the phase space so that 
it is reasonable to introduce  function $D_N(\Phi^I,\Pi_I,t)$ which is function of 
the phase space variables when an expression $D_N(\Phi^I,\Pi_I,t)\prod_{A=1}^N\prod_{B=1}^N d^9\Phi_{AB}d^9\Pi_{AB}$ gives number of members of ensemble that belong to the volume element 
$\prod_{A=1}^N\prod_{B=1}^N d^9\Phi_{AB}d^9\Pi_{AB}$ around the point $\Phi^I_{AB},\Pi_{I,AB}$. It can be shown
(see for example \cite{kintheory}) that $D_N$ is proportional by unspecified constant to $\rho_N(\Phi^I,\Pi_I,t)$ which gives
probability that the system can be found at the point of the phase space $\Phi^I_{AB},\Pi_{I,AB}$ at time $t$. In other words, it determines  probability that the system is in particular microstate.
Note that $\rho_N$ obeys Liouville equation 
\begin{equation}\label{fundLiouv}
	\frac{\partial \rho_N}{\partial t}+\pb{\rho_N,H_N}=0 \ ,
\end{equation}
where
\begin{equation}
	\pb{\rho_N,H_N}=\frac{\partial \rho_N}{\partial \Phi^I_{AB}}\frac{\partial H_N}{\partial \Pi_{I,AB}}
	-\frac{\partial \rho_N}{\partial \Pi_{I,AB}}\frac{\partial H_N}{\partial \Phi^I_{AB}}\ . 
\end{equation}
Let us now consider situation when we are interested in the dynamics of $n-$D0-branes where $n<N$. It is clear that their
configuration is also affected by the rest of the system. 
To see this explicitly  let us write $N\times N$ matrix $\Phi^I$ as
\begin{equation}
	\Phi^I=\left(\begin{array}{cc}
	\phi^I & X^I \\
	Y^I & Z^I \\ \end{array}\right) \ ,
\end{equation}
where $\phi^I$ is $n\times n$ matrix, $X^I$ is $n\times (N-n)$ matrix,
$Y^I$ is $(N-n)\times  n$ matrix and finally $Z^I$ is $(N-n)\times (N-n)
$ matrix. Let us introduce conjugate momenta $\pi_{I,\alpha\beta}$   to $\phi^I_{\alpha\beta}$ where $\alpha,\beta=1,\dots,n$ with non-trivial Poisson bracket
\begin{equation}
	\pb{\phi^I_{\alpha\beta},\pi_{J,\gamma\delta}}=
\delta^I_J\delta_{\alpha \gamma}\delta_{\beta\delta} \ .
\end{equation}
Further, let $\Pi^X_{I,\alpha j},\alpha=1,\dots,n,i=1,\dots,N-n$ are conjugate momenta to $X^I_{\alpha,i}$ with non-zero Poisson brackets
\begin{equation}
	\pb{X^I_{\alpha i},\Pi^X_{J,\beta j}}=\delta^I_J\delta_{\alpha\beta}\delta_{ij} \ . 
\end{equation}
In the same way we deal with the matrix $Y^I_{i,\alpha}$ with corresponding conjugate momenta
$\Pi^Y_{J,j\beta}$
\begin{equation}
	\pb{Y^I_{i\alpha},\Pi^Y_{J,j\beta}}=\delta^I_J\delta_{ij}\delta_{\alpha\beta}
\end{equation}
and finally with $Z^I_{ij}$ and conjugate momentum $\Pi^Z_{J,ij}$
\begin{equation}
\pb{Z^I_{ij},\Pi^Z_{J,kl}}=\delta^I_J\delta_{ik}\delta_{jl} \ . 
\end{equation}
Let us now define $n-$point function for subspace of the phase space spanned by 
$\phi^I_{\alpha,\beta},\pi_{J,\alpha\beta}$ by integrating out all remaining degrees of freedom
in the $N$ correlation function. Before we proceed to the explicit definition of $\rho_n$ let us introduce 
 following notation
\begin{eqnarray}
&&\bphi^I_{\alpha\beta}=(\phi^I_{\alpha\beta},\pi_{\alpha\beta}) \ , \quad d\bphi_{\alpha\beta}=
d^9\phi^I_{\alpha\beta}d^9\pi_{I,\alpha\beta} \ , \nonumber \\
&&\bX^I_{\alpha k}=(X^I_{\alpha k},\Pi^X_{I,\alpha k}) \ , \quad d\bX_{\alpha k}=d^9X_{\alpha k}
d^9\Pi^X_{\alpha k} \ , \nonumber \\
&&\bY^I_{l \beta}=(Y^I_{l\beta},\Pi^Y_{I,l\beta}) \ , \quad d\bY_{l\beta}=d^9Y_{l\beta}d^9\Pi^Y_{l\beta} \ , \nonumber \\
&&\bZ^I_{kl}=(Z^I_{kl},\Pi^Z_{kl}) \ , \quad d\bZ_{kl}=d^9Z_{kl}d^9\Pi^Z_{kl} \ . \nonumber \\
\end{eqnarray}
Let us finally define volume element $d\Omega_n$ as
\begin{equation}\label{defOmegan}
d\Omega_n\equiv  \prod_{\alpha=1}^n	\prod_{k=1}^{N-n} d\bX_{\alpha k}
\prod_{\beta=1}^n\prod_{l=1}^{N-n}	d\bY_{l\beta}
\prod_{r=1}^{N-n}\prod_{s=1}^{N-n}d\bZ_{rs}	\ , 
\end{equation}
where by definition we obtain following important relation between 
$d\Omega_n$ and $d\Omega_{n+1}$
\begin{equation}\label{relOmegan}
	d\Omega_n=\prod_{\alpha=1}^n d\bX_{\alpha 1}
\prod_{\beta=1}^n d^9\bY_{1\beta} d\bZ_{11}	
d\Omega_{n+1} \ .
\end{equation}
Then we are ready to   define reduced distribution function  $\rho_{n}$ as
\begin{equation}
	\rho_n(\bphi)=\int d\Omega_n
	\rho_N(\bphi,\bX,\bY,\bZ) \ . 
\end{equation}
This is natural definition since by integration of $N-$particle
distribution function over complement of subspace of the phase space defined by  $\phi_{\alpha\beta}$ and $\pi_{\alpha\beta}$
we sum over all probabilities which are not relevant for dynamics of $\phi_{\alpha\beta},\pi_{\alpha\beta}$. However clearly the presence of degrees of freedom we integrate over affect the dynamics of reduced distribution function as follows from the 
fundamental equation of motion for $\rho_N$
(\ref{fundLiouv}) when we integrate it  over volume element $d\Omega_n$. First of all we get
\begin{equation}\label{partrhon}
	\int d\Omega_n \frac{\partial \rho_N}{\partial t}=
	\frac{\partial \rho_n}{\partial t} \ 
\end{equation}
so that (\ref{fundLiouv}) implies
\begin{equation}\label{defparrhon}
	\partial_t \rho_n=-\int d\Omega_n\pb{\rho_N,H_N}=-\pb{\rho_n,H_N} \ 
\end{equation}
and our goal is to calculate right side of (\ref{defparrhon}).
In the first step we should split $H_N$ in an appropriate way.  Using  
%
%
%
%
%
%
%
%
\begin{eqnarray}
\tr (\Pi_I\delta^{IJ}\Pi_{J})=
\pi_{I,\alpha\beta}\pi^I_{\beta\alpha}+
\Pi^X_{I,\alpha j}\delta^{IJ}\Pi^Y_{J,j\alpha}+
\Pi^Y_{I,i\beta}\delta^{IJ}\Pi^X_{J,\beta i}+
\Pi^Z_{I,ij}\delta^{IJ}\Pi^Z_{J,ji} \  \nonumber \\
\end{eqnarray}
and also 
\begin{equation}
	[\Phi^I,\Phi^J]=\left(\begin{array}{c|c}
		[\phi^I,\phi^J]+X^IY^J-X^JY^I & \phi^IX^J+X^IZ^J-\phi^JX^I-X^JZ^I \\
		\midrule
		Y^I\phi^J-Y^J\phi^I+Z^IY^J-Z^JY^I & Y^IX^J-Y^JX^I+[Z^I,Z^J] \\ \end{array}
	\right)
\end{equation}
we find that the total Hamiltonian $H_N$ can be written as 
\begin{eqnarray}
H_N=H_n+H_{int}+H_{(N-n)} \ , \nonumber \\
\end{eqnarray}
where
\begin{eqnarray}
&&	H_n=\frac{g_sl_s}{2}
	\pi_{I,\alpha\beta}\pi^I_{\beta\alpha}+
\frac{1}{4g_sl_s}\tr [\phi^I,\phi^J][\phi_I,\phi_J] \ , 	
	\nonumber \\
&&	H_{(N-n)}=\frac{g_sl_s}{2}
\tr_i \Pi^Z_{I}\delta^{IJ}\Pi_{J}^Z+\frac{1}{4g_sl_s}
\tr_i[Z^I,Z^J][Z_I,Z_J] \ , \nonumber \\	
&&H_{int}
=\frac{g_sl_s}{2}(
\Pi^X_{I,\alpha j}\delta^{IJ}\Pi^Y_{J,j\alpha}+
\Pi^Y_{I,i\beta}\delta^{IJ}\Pi^X_{J,\beta i})+
\frac{1}{2g_s l_s}\tr[\phi^I,\phi^J](X_IY_J-X_JY_I)+
\nonumber \\
&&+\frac{1}{2g_sl_s}\tr (X^IY^J-X^JY^I)(X_IY_J-X_JY_I)
+\frac{1}{2g_sl_s}\tr (X^IZ^J-Z^JX_I)(Z_IY_J-Z_JY_I)+
\nonumber \\
&&+\frac{1}{2g_sl_s}\tr(\phi^IX^J-\phi^JX^I)(Y_I\phi_J-Y_J\phi_I)
+\frac{1}{2g_sl_s}\tr (X^IZ^J-X^JZ^I)(Y_I\phi_J-Y_J\phi_I)
+\nonumber \\
&&+\frac{1}{2g_sl_s}\tr(\phi^I X^J-\phi^JX^I)(Z_IY_J-Z_JY_I)+
\frac{1}{2g_sl_s}\tr [Z^I,Z^J](Y_IX_J-Y_JX_I) \ . \nonumber \\
\end{eqnarray}
Now we are ready to proceed to the calculation of  Poisson 
bracket in (\ref{defparrhon}). 
First of all we have
\begin{eqnarray}
&&	\pb{\rho_n,H_{(N-n)}}=
\int d\Omega_n \pb{\rho_N,H_{(N-n)}}=\nonumber \\
&&
=
\int d\Omega_n g_sl_s  \left( \frac{\partial \rho_N}{\partial Z^I_{ij}}
\delta^{IJ}\Pi^Z_{J,ij}-\frac{\partial \rho_N}{\partial \Pi^Z_{I,ij}}\frac{\partial (\dots)}{\partial Z^I_{ij}}\right)=0
\nonumber \\
\end{eqnarray}
since each partial derivative meats corresponding integration and hence its contribution is zero due to the fact that we impose appropriate boundary condition on the function $\rho_N$ that say that $\rho_N$  goes sufficiently quickly to zero in asymptotic regions $Z\rightarrow \pm \infty, \Pi^Z\rightarrow \pm \infty$. 

In case of the Poisson bracket between $\rho_n$ and $H_{int}$ the situation is more involved. 
Firstly we again find that some terms vanish thanks to the integration by parts and imposing appropriate
boundary conditions, for example
\begin{eqnarray}
&&	\pb{\rho_n,\frac{g_s l_s}{2}(\Pi^X_{I,\alpha j}\delta^{ij}
\Pi^Y_{J,j\alpha}+\Pi^Y_{I,i\beta}\delta^{IJ}\Pi^X_{J,\beta i})}=
\nonumber \\
\nonumber \\		
&&=g_sl_s\int d\Omega_n (\frac{	\partial \rho_N}{\partial X^I_{\alpha j}}\delta^{IJ}\Pi^Y_{J,j\alpha}+\frac{\partial \rho_N}{\partial Y^J_{j\alpha}}\delta^{JI}\Pi^X_{I,\alpha j})=0
\nonumber \\
\end{eqnarray}
by integration by parts.  In the same it can be easily shown  that contributions from the potential term in $H_{int}$ that are solely functions of $X,Y,Z$,  vanish. 
On the other hand when we calculate Poisson brackets between $\rho_n$ and terms that depend on $\phi$ we obtain non-zero result. Let us start with following Poisson bracket
\begin{eqnarray}
	\pb{\rho_n,\frac{1}{2g_sl_s}\tr [\phi^I,\phi^J](X_IY_J-X_JY_I)}
=-\frac{1}{g_sl_s}\int d\Omega_n\frac{\partial\rho_N}{\partial \pi_{I,\alpha\beta}}[\phi^J,(X_IY_J-X_JY_I)]_{\beta\alpha}\ , 
\nonumber \\
\end{eqnarray}
where we again ignored boundary terms. In the similar way we get 
\begin{eqnarray}
&&	\pb{\rho_n,
\frac{1}{2g_sl_s}\tr (\phi^IX^J-\phi^JX^I)(Y_I\phi_J-Y_J\phi_I)}=
\nonumber \\
&&=-\frac{1}{g_s l_s}
\int d\Omega_n \frac{\partial \rho_N}{\partial \pi_{I,\alpha\beta}}[
X^K_{\beta l}(Y_I\phi_K-Y_K\phi_I)_{l\alpha}-(\phi^IX^K-\phi^KX^I)_{\beta m}Y_{K,m\alpha}] \ , \nonumber \\
\nonumber \\
\end{eqnarray}
where we again used integration by parts and hence terms proportional to $\frac{\partial \rho_N}{\partial \Pi^{X}},\frac{\partial \rho_N}{\partial \Pi^Y}$ vanish.
Finally there is non-zero contribution 
\begin{eqnarray}
&&	\pb{\rho_n,\frac{1}{2g_sl_s}\tr((\phi^KX^L-\phi^LX^K)(Z_KY_L-Z_LY_K)+(X^KZ^L-X^LZ^K)(Y_K\phi_L-Y_L\phi_K)}=\nonumber \\
&&=\frac{1}{g_sl_s}
\int d\Omega_n\frac{\partial \rho_N}{\partial \pi_{I,\alpha\beta}}
(X^K_{\beta  j}(Z_KY_I-Z_IY_K)_{j\alpha}+
(X^IZ^L-X^LZ^I)_{\beta m}Y_{L,m\alpha}) \ . 
\nonumber \\
\end{eqnarray}
Collecting all terms together we obtain 
\begin{equation}
\partial_t \rho_n+\pb{\rho_n,H_n}=-\pb{\rho_n,H_{int}}=\int d\Omega_n\hat{L}
\rho_N \ , 
\end{equation}
where we defined operator $\hat{L}$ by prescription 
\begin{eqnarray}
&&\int d\Omega_n \hat{L}\rho_N
=\frac{1}{g_sl_s}\int d\Omega_n\frac{\partial\rho_N}{\partial \pi_{I,\alpha\beta}}[\phi^J,(X_IY_J-X_JY_I)]_{\beta\alpha}+
\nonumber \\
&&+\frac{1}{g_s l_s}
\int d\Omega_n \frac{\partial \rho_N}{\partial \pi_{I,\alpha\beta}}[
X^K_{\beta l}Y_I\phi_K-Y_K\phi_I)_{l\alpha}-(\phi^IX^K-\phi^KX^I)_{\beta m}Y_{K,m\alpha}]-\nonumber \\
&&-\frac{1}{g_sl_s}
\int d\Omega_n\frac{\partial \rho_N}{\partial \pi_{I,\alpha\beta}}(
X^K_{\beta  j}(Z_KY_I-Z_IY_K)_{j\alpha}+
(X^IZ^L-X^LZ^I)_{\beta m}Y_{L,m\alpha}) \ .
\nonumber \\
\end{eqnarray}
To proceed further we introduce an important presumption that all 
D0-branes are equivalent and that the partition function $f_N$ is symmetric in its arguments. Schematically 
\begin{equation}
	\rho_N(\bphi,\bX_{\alpha 1},\bX_{\alpha 2},\dots)=
	\rho_N(\bphi,\bX_{\alpha 2},,\bX_{\alpha 1},\dots) \ . 
\end{equation}
In order to demonstrate an importance of this presumption let us consider special case when 
$n=N-2$ and 
 start with the expression 
\begin{eqnarray}
&&\frac{1}{g_s l_s}
\int \prod_{\gamma=1}^{N-2}d\bX_{\gamma 1}
\prod_{\delta=1}^{N-2}d\bX_{\delta 2}
\prod_{\omega=1}^{N-2}d\bY_{1\omega}\prod_{\sigma=1}^{N-2}d\bY_{\sigma 2}
d\bZ_{11}d\bZ_{12}d\bZ_{21}d\bZ_{22}\times 
\nonumber \\
&&\times  \frac{\partial \rho_N}{\partial \pi_{I,\alpha\beta}}
[\phi^J,(X_IY_J-X_JY_I)]_{\beta\alpha} \nonumber \\
\end{eqnarray}
and consider first term in the commutator
\begin{eqnarray}
&&\frac{1}{g_s l_s}
	\int \prod_{\gamma=1}^{N-2}d\bX_{\gamma 1}
	\prod_{\delta=1}^{N-2}d\bX_{\delta 2}
	\prod_{\omega=1}^{N-2}d\bY_{1\omega}\prod_{\sigma=1}^{N-2}d\bY_{\sigma 2}
	d\bZ_{11}d\bZ_{12}d\bZ_{21}d\bZ_{22}\times 
	\nonumber \\
&&	\times  \frac{\partial \rho_N(\bphi,\bX_{\alpha 1},\bY_{\beta 1},\bX_{\gamma 2},\bY_{\gamma 2}.\dots)}{\partial \pi_{I,\alpha\beta}}
\phi^J_{\beta\gamma}(X_{I,\gamma 1}Y_{J,1\alpha}+X_{I,\gamma 2}Y_{J,2\alpha})= \nonumber \\
&&\frac{1}{g_s l_s}
\int \prod_{\gamma=1}^{N-2}d\bX_{\gamma 1}
\prod_{\delta=1}^{N-2}d\bX_{\delta 2}
\prod_{\omega=1}^{N-2}d\bY_{1\omega}\prod_{\sigma=1}^{N-2}d\bY_{\sigma 2}
d\bZ_{11}d\bZ_{12}d\bZ_{21}d\bZ_{22}\times 
\nonumber \\
&&\times  \frac{\partial \rho_N(\bphi,\bX_{\alpha 1},\bY_{\beta 1},\bX_{\gamma 2},\bY_{\gamma 2}.\dots)}{\partial \pi_{I,\alpha\beta}}
\phi^J_{\beta\gamma}(X_{I,\gamma 1}Y_{J,1\alpha}+\nonumber \\	
&&+\frac{1}{g_s l_s}
\int \prod_{\gamma=1}^{N-2}d\bX_{\gamma 1}
\prod_{\delta=1}^{N-2}d\bX_{\delta 2}
\prod_{\omega=1}^{N-2}d\bY_{1\omega}\prod_{\sigma=1}^{N-2}d\bY_{\sigma 2}
d\bZ_{11}d\bZ_{12}d\bZ_{21}d\bZ_{22}\times 
\nonumber \\
&&\times  \frac{\partial \rho_N(\bphi,\bX_{\alpha 2},\bY_{\beta 2},\bX_{\gamma 1},\bY_{\gamma 1}.\dots)}{\partial \pi_{I,\alpha\beta}}
\phi^J_{\beta\gamma}X_{I,\gamma 2}Y_{J,2\alpha} \ , \nonumber \\	
\end{eqnarray}
where in the last step we used symmetry presumption 
$\rho_N(\dots,\bX_{\alpha 1},\bX_{\alpha 2},\bY_{1\beta },\bY_{2 \beta},\dots)=
\rho_N(\dots,\bX_{\alpha 2},\bX_{\alpha 1},\bY_{2\beta },\bY_{1 \beta},\dots)$. If we now perform  relabeling variables as
$\bX_{\alpha 2}\rightarrow \bX_{\alpha 1}, \bY_{2\beta}\rightarrow \bY_{1\beta}$ which can be easily done since they are integration variables we find that the expression above can be written as
\begin{eqnarray}
&&\frac{1}{g_s l_s}
	\int \prod_{\gamma=1}^{N-2}d\bX_{\gamma 1}
	\prod_{\delta=1}^{N-2}d\bX_{\delta 2}
	\prod_{\omega=1}^{N-2}d\bY_{1\omega}\prod_{\sigma=1}^{N-2}d\bY_{\sigma 2}
	d\bZ_{11}d\bZ_{12}d\bZ_{21}d\bZ_{22}\times 
	\nonumber \\
&&	\times  \frac{\partial \rho_N(\bphi,\bX_{\alpha 1},\bY_{\beta 1},\bX_{\gamma 2},\bY_{\gamma 2}.\dots)}{\partial \pi_{I,\alpha\beta}}
\phi^J_{\beta\gamma}(X_{I,\gamma 1}Y_{J,1\alpha}+X_{I,\gamma 2}Y_{J,2\alpha})= \nonumber \\
&&=\frac{2}{g_sl_s}\int \prod_{\gamma=1}^{N-2}
d\bX_{\gamma 1}\prod_{\delta=1}^{N-2}d\bY_{1\delta}
d\bZ_{11}\frac{\partial \rho_{N-1}}{\partial \pi_{I,\alpha\beta}}
\phi^J_{\beta\gamma}X_{I,\gamma 1}Y_{J,1\alpha} \ , \nonumber \\
\end{eqnarray}
where we introduced $\rho_{N-1}$ defined as 
\begin{equation}
	\rho_{N-1}(\bphi,\bX_{\alpha 1},\bY_{1\beta},\bZ_{11})=
	\int \prod_{\gamma=1}^{N-2}d\bX_{\gamma 2}\prod_{\delta=1}^{N-2}d\bY_{2\delta}
	d\bZ_{12}d\bZ_{21}d\bZ_{11}\rho_N \ . 
\end{equation}
In the same way we can proceed with remaining terms and we obtain the result
\begin{eqnarray}
&&\frac{1}{g_s l_s}
\int \prod_{\gamma=1}^{N-2}d\bX_{\gamma 1}
\prod_{\delta=1}^{N-2}d\bX_{\delta 2}
\prod_{\omega=1}^{N-2}d\bY_{1\omega}\prod_{\sigma=1}^{N-2}d\bY_{\sigma 2}
d\bZ_{11}d\bZ_{12}d\bZ_{21}d\bZ_{22}\times 
\nonumber \\
&&\times  (\frac{\partial \rho_N}{\partial \pi_{I,\alpha\beta}}
{\partial \pi_{I,\alpha\beta}}[\phi^J,(X_IY_J-X_JY_I)]_{\beta\alpha}= \nonumber \\
&&=\frac{2}{g_sl_s}\int \prod_{\gamma=1}^{N-2}
d\bX_{\gamma 1}\prod_{\delta=1}^{N-2}d\bY_{1\delta}
d\bZ_{11}	\times \nonumber \\
&&\times \frac{\partial \rho_{N-1}}{\partial \pi_{I,\alpha\beta}}
(\phi^J_{\beta\gamma}(X_{I,\gamma 1}Y_{J,1\alpha}-X_{J,\gamma 1}
Y_{I,1\alpha})-(X_{I,\beta 1}Y_{J,1\gamma}-X_{J,\beta 1}
Y_{I,1\gamma})\phi^J_{\gamma \alpha}) \ . 
 \nonumber \\
\end{eqnarray}
Clearly this expression can be generalized to the $n-$point function and we get 
\begin{eqnarray}
&&\frac{1}{g_sl_s}\int d\Omega_n\frac{\partial\rho_N}{\partial \pi_{I,\alpha\beta}}[\phi^J,(X_IY_J-X_JY_I)]_{\beta\alpha}
=\frac{(N-n)}{g_sl_s}\int \prod_{\gamma=1}^{n}
d\bX_{\gamma 1}\prod_{\delta=1}^{n}d\bY_{1\delta}
d\bZ_{11}	\times \nonumber \\
&&\times \frac{\partial \rho_{n+1}(\bphi,\bX_{\alpha 1},\bY_{1\beta},\bZ_{11})}{\partial \pi_{I,\alpha\beta}}
(\phi^J_{\beta\gamma}(X_{I,\gamma 1}Y_{J,1\alpha}-X_{J,\gamma 1}
Y_{I,1\alpha})-(X_{I,\beta 1}Y_{J,1\gamma}-X_{J,\beta 1}
Y_{I,1\gamma})\phi^J_{\gamma \alpha}) \ . 
\nonumber \\
	\end{eqnarray}
As the next step we consider an expression 
\begin{equation}
-\frac{1}{g_sl_s}
\int d\Omega_n\frac{\partial \rho_N}{\partial \pi_{I,\alpha\beta}}(
X^K_{\beta  j}(Z_KY_I-Z_IY_K)_{j\alpha}+
+(X^IZ^L-X^LZ^I)_{\beta m}Y_{L,m\alpha})
\end{equation}
and again consider $n=N-2$ function and restrict to the first term in the sum give
above 
\begin{eqnarray}
&&-\frac{1}{g_s l_s}
	\int \prod_{\gamma=1}^{N-2}d\bX_{\gamma 1}
	\prod_{\delta=1}^{N-2}d\bX_{\delta 2}
	\prod_{\omega=1}^{N-2}d\bY_{1\omega}\prod_{\sigma=1}^{N-2}d\bY_{\sigma 2}
	d\bZ_{11}d\bZ_{12}d\bZ_{21}d\bZ_{22}\times 
	\nonumber \\
&&	\times  \frac{\partial \rho_N(\bphi,\bX_{\alpha 1},\bX_{\alpha 2},
		\bY_{1\beta},\bY_{2\beta},\bZ_{11},\bZ_{12},\bZ_{21},\bZ_{22})}{\partial \pi_{I,\alpha\beta}}\times
	\nonumber \\
	\times 
&&(X^K_{\beta 1}Z_{K,11}Y_{I,1\alpha}+X^K_{\beta 1}Z_{K,12}Y_{I,2\alpha}+
X^K_{\beta 2}Z_{K,21}Y_{I,1\alpha}+X^K_{\beta 2}Z_{K,22}Y_{I,2\alpha})=
\nonumber \\		
&&=-\frac{4}{g_sl_s}
\int \prod_{\gamma=1}^{N-2}d\bX_{\gamma 1}
\prod_{\omega=1}^{N-2}d\bY_{1\omega}
d\bZ_{11} \frac{\partial \rho_{N-1}(\bphi,\bX_{\alpha 1},\bX_{\alpha 2},\bZ_{11})}
{\partial \pi_{I,\alpha\beta}}X^K_{\beta 1}Z_{K,11}Y_{I,1\alpha} \ .\nonumber \\
\end{eqnarray}
In the same way we can proceed with remaining terms in the sum and generalize it to  arbitrary $n$ with the result
\begin{eqnarray}
&& -	\frac{1}{g_sl_s}
	\int d\Omega_n\frac{\partial \rho_N}{\partial \pi_{I,\alpha\beta}}(
	X^K_{\beta  j}(Z_KY_I-Z_IY_K)_{j\alpha}+
	+(X^IZ^L-X^LZ^I)_{\beta m}Y_{L,m\alpha})=
	\nonumber \\
&&=-\frac{(N-n)^2}{g_sl_s}
\int \prod_{\gamma=1}^n d\bX_{\gamma 1}\prod_{\delta=1}^n
d\bY_{1\beta}d\bZ_{11}\frac{\partial \rho_{n+1}}{\partial \pi_{I,\alpha\beta}}
\times \nonumber \\
&&\times (X^K_{\beta 1}(Z_{K,11}Y_{I,1\alpha}-Z_{I,11}Y_{K,1\alpha})+
(X^I_{\beta 1}Z^K_{11}-X^K_{\beta 1}Z^I_{11})Y_{K,1\alpha}) \ . 
\nonumber \\
\end{eqnarray}
This procedure can be repeated with remaining terms in the definition of 
$\hat{L}$ so that we get a sequence of the partial differential equation 
for partition function $\rho_n$ that has the form of famous BBGKY hierarchy
\begin{eqnarray}
&&	\partial_t \rho_n+\pb{\rho_n,H_n}=\int 
\prod_{\gamma=1}^n d\bX_{\gamma 1}\prod_{\delta=1}^n d\bY_{1\delta}
d\bZ_{11} 	\hat{L}\rho_{n+1}(\bphi,\bX_{\alpha 1},\bY_{1\beta},\bZ_{11}) \ , 
\nonumber \\
&&\hat{L}\rho_{n+1}=
-\frac{(N-n)^2}{g_sl_s}
\frac{\partial \rho_{n+1}}{\partial \pi_{I,\alpha\beta}}
\times \nonumber \\
&&\times (X^K_{\beta 1}(Z_{K,11}Y_{I,1\alpha}-Z_{I,11}Y_{K,1\alpha})+
(X^I_{\beta 1}Z^K_{11}-X^K_{\beta 1}Z^I_{11})Y_{K,1\alpha})
+\nonumber \\
&&+\frac{(N-n)}{g_sl_s} \frac{\partial \rho_{n+1}}{\partial \pi_{I,\alpha\beta}}[
X^K_{\beta l}(Y_{I,1\delta}\phi_{K,\delta\alpha}-Y_{K,1\delta}\phi_{I,\delta\alpha})-(\phi^I_{\beta\gamma}X^K_{\gamma 1}-\phi^K_{\beta \gamma}X^I_{\gamma 1})Y_{K,1\alpha}]+
\nonumber \\
&&+\frac{(N-n)}{g_sl_s} \frac{\partial \rho_{n+1}}{\partial \pi_{I,\alpha\beta}}
(\phi^J_{\beta\gamma}(X_{I,\gamma 1}Y_{J,1\alpha}-X_{J,\gamma 1}
Y_{I,1\alpha})-(X_{I,\beta 1}Y_{J,1\gamma}-X_{J,\beta 1}
Y_{I,1\gamma})\phi^J_{\gamma \alpha}) \ . 
\nonumber \\
\end{eqnarray}
Note that this is an exact sequence of $N$ partial differential equations for $\rho_n$. An exceptional equation corresponds to $n=1$ that has the form 
\begin{eqnarray}\label{rho1}
&&	\partial_t \rho_1+\pb{\rho_1,H_1}=\int 
	 d\bX d\bY
	d\bZ 	\hat{L}\rho_{2}(\bphi,\bX,\bY,\bZ) \ , 
	\nonumber \\
&&	\hat{L}\rho_{2}=-\frac{(N-1)^2}{g_sl_s}\frac{\partial \rho_{2}}{\partial \pi_I}[	X^K(Z_{K}Y_{I}-Z_{I}Y_{K})
	+(X^IZ^K-X^KZ^I)Y_K]
	+\nonumber \\
&&	+\frac{(N-1)}{g_sl_s} \frac{\partial \rho_{2}}{\partial \pi_{I}}[
	X^K(Y_{I}\phi_{K}-Y_{K}\phi_{I})-(\phi^IX^K-\phi^KX^I)Y_{K}] \ , 	\quad 
	H_1=\frac{g_sl_s}{2}\pi_I\pi^I \ , \nonumber \\
	\nonumber \\
\end{eqnarray}
where we defined
\begin{equation}
	\bX\equiv \bX_{11} \ , \quad  \bY=\bY_{11} \ , \quad \bZ\equiv \bZ_{11} \ . 
\end{equation}
From (\ref{rho1}) we would get kinetic equation for D0-brane when 
we were able to express $\rho_2$ as function of $\rho_1$. Note that in case of the Boltzmann kinetic equation the right side
of the kinetic equation contains collision term
that can be derived by reasonable arguments from one point function in BBGKY hierarchy (\cite{kintheory}). 
 It would be extremely interesting 
to develop similar approximation in case of the system of D0-branes which is now under investigation.
\\
\\
 {\bf Acknowledgment:}
This work  was
supported by the Grant Agency of the Czech Republic under the grant
GA26-22343S.

	\end{document}